\documentclass[12pt,showpacs,showkeys]{revtex4}

\usepackage{amssymb,amsmath}

\usepackage{epsf,epsfig}

\def\a{\alpha}

\begin{document}

\title{Measuring a piecewise constant axion field in classical
  electrodynamics}

\author{Yuri N.~Obukhov\footnote{On leave from: Dept. of Theoret.
    Physics, Moscow State University, 117234 Moscow, Russia} and
  Friedrich W.~Hehl\footnote{Also at: Dept.\ of Physics \& Astron.,
    University of Missouri-Columbia, Columbia, MO 65211, USA}}
\affiliation{Institute for Theoretical Physics, University of Cologne,
  50923 K\"oln, Germany}
\bigskip


\begin{abstract}
  In order to settle the problem of the ``Post constraint" in material
  media, we consider the propagation of a plane electromagnetic wave
  in a medium with a piecewise constant axion field. Although a
  constant axion field does not affect the wave propagation in a
  homogeneous medium, we show that the reflection and transmission of
  a wave at an interface between the two media is sensitive to the
  difference of the axion values. This observation can be used to
  determine experimentally the axion piece in matter despite the fact
  that a constant axion value does not contribute to the Maxwell
  equations.
\end{abstract}

\pacs{03.50.De, 46.05.+b, 14.80.Mz} 

\keywords{axion piece, material media, magnetoelectric effect,
  measurement of axion piece, axion electrodynamics} 


\maketitle

\section{Local and linear media with axion piece}

Recently, we discussed local and linear media in classical
electrodynamics \cite{postcon}. In particular, we investigated
possible magnetoelectric effects, which are related to the crossterms
between the magnetic (electric) field strength and the electric
(magnetic) excitation. Using the premetric formalism of
electrodynamics, see \cite{Birkbook,postcon,Zlatibor}, the most
general local and linear constitutive relation can be written as
\begin{eqnarray}\label{explicit'}
  {\cal D}^a\!&=\!&\left( \varepsilon^{ab}\hspace{4pt} - \,
    \epsilon^{abc}\,n_c \right)E_b\,+\left(\hspace{9pt} \gamma^a{}_b +
    s_b{}^a - \delta_b^a s_c{}^c\right) {B}^b + \alpha\,B^a \,,\\
  {\cal H}_a\! &=\! &\left( \mu_{ab}^{-1} - \hat{\epsilon}_{abc}m^c
  \right) {B}^b +\left(- \gamma^b{}_a + s_a{}^b - \delta_a^b
    s_c{}^c\right)E_b - \alpha\,E_a\,.\label{explicit''}
\end{eqnarray}
Here $ {\cal D}^a$ and ${\cal H}_a$ are the electric and magnetic
excitations, respectively, and $E_a$ and $ {B}^a$ the electric and the
magnetic field strengths. The Kronecker symbol is denoted by
$\delta_a^b\,$, the totally antisymmetric Levi-Civita symbol by
$\epsilon^{abc}$ and $\hat{\epsilon}_{abc}$, respectively. We have 36
constitutive functions or moduli: the permittivity matrix
$\varepsilon^{ab}=\varepsilon^{ba}$ (6 independent components), the
impermeability matrix $\mu^{-1}_{ab}=\mu^{-1}_{ba}$ (6 components),
the tracefree principal magnetoelectric matrix $\gamma^a{}_b$ (8
components), the 15 skewon pieces $m^a,n_a,s_a{}^b$, and, eventually,
1 axion piece $\alpha$.  Such a local and linear medium with 36 moduli
--- called sometimes bi-anisotropic --- has been considered, amongst
others, by Lindell, Sihvola, Tretyakov, and collaborators
\cite{Ismo,Lindell1994,SihvolaLindell1995,Tretyakov1998}.

In conventional materials the {\it skewon\/} and the {\it axion\/}
pieces {\it vanish\/} and we are left with
\begin{eqnarray}\label{convent1}
  {\cal D}^a&=&\hspace{2pt}\varepsilon^{ab}\,E_b+ \gamma^a{}_b
  \,B^a\,,\\ {\cal H}_a&=& \mu_{ab}^{-1}
  {B}^b-\gamma^b{}_a\,E_b\,.\label{convent2}
\end{eqnarray}
The principal magnetoelectric cross terms induced by $\gamma^a{}_b$
are known to exist in various media, see O'Dell \cite{O'Dell}. Such
media are described by $6+6+8=20$ constitutive functions.
Eqs.(\ref{convent1}) and (\ref{convent2}) represent the most general
material considered by Post \cite{Post}, e.g..

The skewon pieces are not considered in this article, see, however,
\cite{skewon}. We address here the question on whether the axion piece
is permitted --- thereby possibly extending a material with 20 moduli
to one with 21 moduli --- and if so whether it can be determined
experimentally by standard methods. In Post \cite{Post} it was argued
that the axion piece has to vanish, i.e., $\alpha=0$ (and even
$d\alpha=0$). For this reason, Lakhtakia
\cite{Akhlesh2004a,Akhlesh2004b} (and references given there) called
$\a=0$ the Post constraint and advocated it as a condition each medium
has to fulfill. We quoted in \cite{postcon} literature in which
materials are described (Cr$_2$O$_3$ and Fe$_2$TeO$_6$) that carry an
axion piece. Moreover, Lakhtakia, loc.cit., pointed out that a
constant axion piece $\alpha$ should not be measurable since it drops
out of the Maxwell equations.  However, this is only true if we have
an axion piece that is {\it globally constant\/} at {\it all\/}
spatial points at {\it all\/} times. We will show explicitly that a
material with a piecewise constant $\a$ can very well be investigated
experimentally and thereby $\a$ measured uniquely. Thus, neither can
the Post constraint be uphold \cite{postcon} nor poses the
measurability of $\a$ a problem, as we will show below.

\begin{figure}
\epsfxsize=10cm
\epsfbox{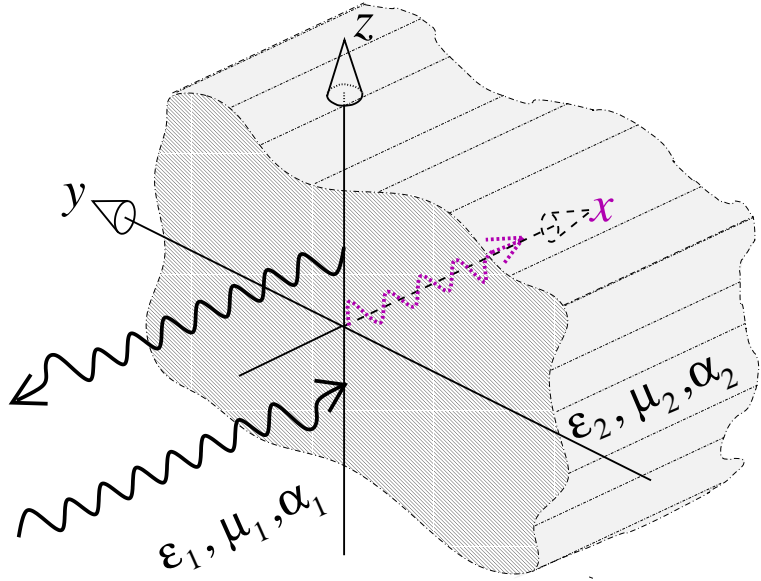}
\caption[{\it Two homogeneous media are separated by the plane $x=0$. Here
  $(x,y,z)$ are Cartesian coordinates. Moreover: Permittivity
  $\varepsilon$, permeability $\mu$, and axion piece $\a$, see
  Eqs.(\ref{con2}) and (\ref{con1}).}] {\label{media}{\it Two
    homogeneous media are separated by the plane $x=0$. Here $(x,y,z)$
    are Cartesian coordinates. Moreover: Permittivity $\varepsilon$,
    permeability $\mu$, and axion piece $\a$, see Eqs.(\ref{con2}) and
    (\ref{con1}).}}
\end{figure}

Accordingly, we discuss here the case of two neighboring homogeneous
media with different but constant axion pieces. They are separated by
the plane $x=0$, as shown in Fig.1. Permittivity and permeability are
assumed to be isotropic. Thus, the constitutive relations for the two
half-spaces carry 3 constitutive constants $\varepsilon, \,\mu^{-1}$,
and $\a$, respectively,
\begin{eqnarray} 
  {\cal D} &=&
(\varepsilon\varepsilon_0)\,\,^{\underline{\star}}E +
  \alpha\,{B}\,,\label{con2} \\{\cal H} &=&
  (\mu\mu_0)^{-1}\>^{\underline{\star}}B - \alpha\,E \,,\label{con1}
\end{eqnarray}
with $^{\underline{\star}}$ as the 3-dimensional Hodge star operator
and $\varepsilon_0$ and $\mu_0$ as electric and magnetic constants (of
the vacuum).  We use here the calculus of differential forms, see
\cite{Birkbook,Ismo}.

As sideremark let us remind ourselves that (\ref{con2}), (\ref{con1}),
formulated 4-dimensionally with the excitation 2-form $H=({\cal
  H},{\cal D})$ and the field strength 2-form $F=(E,B)$ and {\it in
  vacuum,} reduce to {axion\/} (Maxwell-Lorentz) {electrodynamics}
(see Ni \cite{Ni73,Ni77,Ni84} and Wilczek \cite{Wilczek87}) with the
constitutive relation
\begin{equation}
H=\sqrt{\frac{\varepsilon_0}{\mu_0}}\,^\star F+\alpha F\label{vac}\,.
\end{equation}
Here $^\star$ is the 4-dimensional Hodge star operator defined in
terms of the metric of spacetime.  The field equations read
\begin{equation}\label{MaxAx}
  \sqrt{\frac{\varepsilon_0}{\mu_0}}\,d\,^\star F+(d\alpha)\wedge
  F=J\,,\qquad dF=0\,.
\end{equation} 
It is as if the current 3-form $J$ picked up an additional piece
depending on the gradient of the axion field.  For
$\sqrt{{\varepsilon_0}/{\mu_0}}=0$, this corresponds to the pure axion
case, that is, to the perfect electromagnetic conductor (PEMC) of
Lindell and Sihvola
\cite{PEMC1,PEMC2,PEMC3a,PEMC3b,PEMC4,PEMC5,PEMC6}, a structure that
is equivalent to the Tellegen gyrator
\cite{Tellegen1948,Tellegen1956/7}, see also \cite{postcon}. The real
part of Kiehn's {\it chiral vacuum\/} (see \cite{Kiehn2002} and
\cite{Kiehn2004}, pp.\ 140/141) is a subcase, for $\alpha={\rm
  const}$, of axion electrodynamics.

Thus we see that if the axion piece is globally constant, $\alpha
=$const, it does not contribute to the Maxwell equations
(\ref{MaxAx}), even though it emerges in the constitutive relation
(\ref{vac}) and in the boundary conditions to be discussed in the next
section.

\section{Wave propagation in two homogeneous media}

The Maxwell equations without charges and currents read
\begin{eqnarray}\label{Max1}
  \underline{d}{\cal D}& =& 0\,,\qquad \underline{d}{\cal H} -
  \dot{\cal D} = 0\,,\\ \underline{d}B &=& 0\,,\qquad \underline{d}E +
  \dot{B} = 0\,.\label{Max2}
\end{eqnarray}
The calculus of differential forms is used and the conventions of
\cite{Birkbook}. In the absence of the surface charges and currents,
the jump conditions on the boundary surface read (see \cite{Birkbook},
pp.\ 150,151):
\begin{eqnarray}
&&\left({\cal D}_{(2)} - {\cal D}_{(1)}\right)\,\vline\,
{\hbox{\raisebox{-1.5ex}{\scriptsize{S}}}}\wedge \nu = 0,\qquad
\tau_{A}\rfloor\left({\cal H}_{(2)} - {\cal H}_{(1)}\right)\,\vline\,
{\hbox{\raisebox{-1.5ex}{\scriptsize{S}}}}= 0,\label{jump1}\\
&&\left({B}_{(2)} - {B}_{(1)}\right)\,\vline\,
{\hbox{\raisebox{-1.5ex}{\scriptsize{S}}}}\wedge \nu  = 0,\qquad\hspace{4pt}
\tau_{A}\rfloor\left({E}_{(2)} - {E}_{(1)}\right)\,\vline\,
{\hbox{\raisebox{-1.5ex}{\scriptsize{S}}}} = 0.\label{jump2}
\end{eqnarray}
Here $\nu$ is the 1-form density normal to the surface $S$ and
$\tau_{A}$, $A =1,2$, are the two vectors tangential to $S$.  The
constitutive relations were formulated in (\ref{con2}) and
(\ref{con1}).

Let $S$ be the plane $x=0$ in Fig.1 that divides the two parts of
space that are filled with two different homogeneous material media.
For the left half-space $x < 0$, we assume homogeneous matter
characterized by constant values $\varepsilon_1, \mu_1, \alpha_1$.
Similarly, for the right half-space $x > 0$, we have the constant
values $\varepsilon_2, \mu_2, \alpha_2$. Somewhat related situations
with reflected and scattered waves were discussed by Lindell and
Sihvola \cite{PEMC1,PEMC4,PEMC5,PEMC6}.

Consider a plane electromagnetic wave travelling along the $x$ axis in
the left half-space. At the interface $S$, such an incident wave will
be partly reflected and partly refracted into the right half-space.
Accordingly, the ansatz for the electromagnetic field configuration in
the first medium will be a superposition of the right- and left-moving
plane waves. Let $W_y = W_y(\xi_1)$ and $W_z = W_z(\xi_1)$ be the
components of the incident wave with the argument $\xi_1 := \omega t -
k_1x$, whereas $R_y = R_y(\eta)$ and $R_z = R_z(\eta)$ are those of
the reflected wave with the argument $\eta := \omega t + k_1x$. Then,
with the 1-forms $dx$, $dy$, and $dz$,
\begin{eqnarray}
  E &=& \left(W_y + R_y\right)dy + \left(W_z +
    R_z\right)dz,\label{E1}\\B &=& {\frac {k_1}
    \omega}\,dx\wedge\left[\left(W_y - R_y\right)dy + \left(W_z -
      R_z\right)dz\right],\label{B1}\\{\cal D} &=&
  \varepsilon_1\varepsilon_0\left[\left(W_y + R_y\right)dz - \left(W_z
      + R_z\right)dy\right]\wedge dx +\alpha_1\,{\frac {k_1}
    \omega}\,dx\wedge\left[\left(W_y - R_y\right)dy + \left(W_z -
      R_z\right)dz\right],\hspace{20pt}\label{D1}\\{\cal H} &=&
  \varepsilon_1\varepsilon_0{\frac \omega {k_1}}\left[\left(W_y -
      R_y\right)dz - \left(W_z - R_z\right)dy\right] - \alpha_1
  \left[\left(W_y + R_y\right)dy + \left(W_z +
      R_z\right)dz\right].\label{H1}
\end{eqnarray}
A direct check shows that the Maxwell equations
(\ref{Max1}),(\ref{Max2}), together with the constitutive relations
(\ref{con2}),(\ref{con1}), are satisfied provided the only nonvanishing
component of the wave covector has the value
\begin{equation}
{k_1} = {\frac {\omega n_1} c},\qquad {\rm with}\qquad n_1 := 
\sqrt{\varepsilon_1\mu_1}.\label{k1}
\end{equation}

Analogously, the electromagnetic field configuration in the second
half-space is represented by the right-moving transmitted wave: 
\begin{eqnarray}
E &=& T_y\,dy + T_z\,dz,\label{E2}\\
B &=& {\frac {k_2} \omega}\,dx\wedge\left(T_y\,dy + T_z\,dz\right),\label{B2}\\
{\cal D} &=& \varepsilon_2\varepsilon_0\left(T_y\,dz - T_z\,dy\right)\wedge dx 
+\alpha_2\,{\frac {k_2} \omega}\,dx\wedge\left(T_y\,dy + T_z\,dz\right),
\label{D2}\\
{\cal H} &=& \varepsilon_2\varepsilon_0{\frac \omega {k_2}}\left(T_y\,dz 
- T_z\,dy\right) - \alpha_2\left(T_y\,dy + T_z\,dz\right).\label{H2}
\end{eqnarray}
Here the functions $T_y = T_y(\xi_2)$ and $T_z = T_z(\xi_2)$, with the 
argument $\xi_2 := \omega t - k_2x$, describe the transmitted wave in the 
second medium. Analogously to (\ref{k1}) we have 
\begin{equation}
  {k_2} = {\frac {\omega n_2} c},\qquad {\rm with}\qquad n_2 :=
  \sqrt{\varepsilon_2\mu_2}.\label{k2}
\end{equation}

\section{Harmonic waves}

For concreteness, we confine our attention to harmonic waves. Then,
\begin{eqnarray}
W_y(\xi_1)& =& a_1\cos\xi_1 + a_2\sin\xi_1,\qquad  
W_z(\xi_1) = b_1\cos\xi_1 + b_2\sin\xi_1,\\
R_y(\eta)&=& c_1\cos\eta + c_2\sin\eta,\qquad 
R_z(\eta) = d_1\cos\eta + d_2\sin\eta,\\
T_y(\xi_2)& =& p_1\cos\xi_2 + p_2\sin\xi_2,\qquad  
T_z(\xi_2) = q_1\cos\xi_2 + q_2\sin\xi_2.\label{T1}
\end{eqnarray}
In order to construct the complete solution in the two regions, we
have to match the configurations (\ref{E1})-(\ref{H1}) and
(\ref{E2})-(\ref{H2}) on the interface $S$. Using the jump conditions
(\ref{jump1}) and (\ref{jump2}) with $\nu = dx$ and $\tau_A =
(\partial_y,\partial_z)$, we find
\begin{eqnarray}
  \left(W_y +
    R_y\right)\vline\,{\hbox{\raisebox{-1.5ex}{\scriptsize{x=0}}}} &=&
  T_y\vline\,{\hbox{\raisebox{-1.5ex}{\scriptsize{x=0}}}},\label{match1}\\ 
  \left(W_z +
    R_z\right)\vline\,{\hbox{\raisebox{-1.5ex}{\scriptsize{x=0}}}} &=&
  T_z\vline\,{\hbox{\raisebox{-1.5ex}{\scriptsize{x=0}}}},\label{match2}\\ 
  \varepsilon_0c\sqrt{\frac {\varepsilon_1}{\mu_1}}\left(W_y -
    R_y\right){} \vline\,{\hbox{\raisebox{-1.5ex}{\scriptsize{x=0}}}}
  &=& \varepsilon_0c \sqrt{\frac
    {\varepsilon_2}{\mu_2}}\,T_y{}\vline\,{\hbox{\raisebox{-1.5ex}
      {\scriptsize{x=0}}}} -
  [\alpha]\,T_z{}\vline\,{\hbox{\raisebox{-1.5ex}
      {\scriptsize{x=0}}}},\label{match3}\\ \varepsilon_0c\sqrt{\frac
    {\varepsilon_1}{\mu_1}}\left(W_z - R_z\right){}
  \vline\,{\hbox{\raisebox{-1.5ex}{\scriptsize{x=0}}}} &=&
  \varepsilon_0c \sqrt{\frac
    {\varepsilon_2}{\mu_2}}\,T_z{}\vline\,{\hbox{\raisebox{-1.5ex}
      {\scriptsize{x=0}}}} +
  [\alpha]\,T_y{}\vline\,{\hbox{\raisebox{-1.5ex}
      {\scriptsize{x=0}}}}.\label{match4}
\end{eqnarray}
Here $[\alpha]:= \alpha_2 - \alpha_1$ is the {\it jump\/} of the axion
field on the interface $S$. 

The algebraic system (\ref{match1}) to (\ref{match4}) can be
straightforwardly solved. It yields the coefficients of the
reflected ($c_{1,2}$ and $d_{1,2}$) and transmitted ($p_{1,2}$ and
$q_{1,2}$) waves as combinations of those of the incident wave:
\begin{eqnarray}
  c_{1,2} &=& \frac 1{\Delta_\bot}\left[ \left({\frac {\varepsilon_1}
        {\mu_1}} - {\frac {\varepsilon_2}{\mu_2}} - {\frac
        {[\alpha]^2}{\lambda_0^2}}\right)a_{1,2} + 2{\frac {[\alpha]}
      {\lambda_0}}\sqrt{\frac
      {\varepsilon_1}{\mu_1}}\,b_{1,2}\right]\,,\\ d_{1,2} &=& \frac
  1{\Delta_\bot}\left[ -\,2{\frac {[\alpha]}{\lambda_0}} \sqrt{\frac
      {\varepsilon_1}{\mu_1}}\,a_{1,2} + \left({\frac {\varepsilon_1}
        {\mu_1}} - {\frac {\varepsilon_2}{\mu_2}} - {\frac
        {[\alpha]^2}{\lambda_0^2}} \right)\,b_{1,2}\right],\\ p_{1,2}
  &=&
  \frac{2}{\Delta_\bot}\sqrt{\frac{\varepsilon_1}{\mu_1}}\left[\left(
      \sqrt{\frac {\varepsilon_1} {\mu_1}} + \sqrt{\frac
        {\varepsilon_2}{\mu_2}} \right) a_{1,2} + {\frac
      {[\alpha]}{\lambda_0}}\,b_{1,2}\right],\\ q_{1,2} &=&
  \frac{2}{\Delta_\bot}\sqrt{\frac{\varepsilon_1}{\mu_1}}\left[
    -\,{\frac {[\alpha]}{\lambda_0}}\,a_{1,2} + \left(\sqrt{\frac
        {\varepsilon_1} {\mu_1}} + \sqrt{\frac
        {\varepsilon_2}{\mu_2}}\right) b_{1,2}\right].
\end{eqnarray}
Here $\Delta_\bot:= \left(\sqrt{\varepsilon_1/\mu_1} + \sqrt{
    \varepsilon_2/\mu_2}\,\right)^2 + [\alpha]^2/\lambda_0^2\,$. As a
consistency check, we can easily see that the above formulas, for
$[\alpha] = 0$ (that is, either the axion is trivial everywhere or it
has equal values for both material media), reduce to the well known
expressions of the corresponding reflection and transmission
coefficients for a plane wave, see Born and Wolf \cite{BornWolf},
Sec.1.5.

The result obtained clearly shows that despite the fact the constant
axion drops out from the Maxwell field equation, the electromagnetic
wave ``feels" the presence of the axion by experiencing specific
reflection and transmission effects. 

\section{Vacuum with and without an axion piece}

\begin{figure}
\epsfxsize=5truecm
\epsfbox{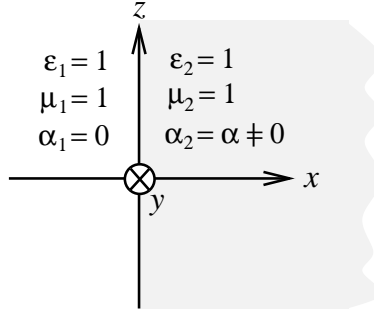}
\caption[{\it On the left-hand side we have pure vacuum and on the 
  right-hand side vacuum together with a constant axion piece $\a$.}]
{\label{fig2}{\it On the left-hand side we have pure vacuum and on the
    right-hand side vacuum together with a constant axion piece
    $\a$.}}
\end{figure}

In order to discuss the measureability of $\a$, let us consider the
case when the first region of the space is vacuum, with $\varepsilon_1
= \mu_1 =1$ and $\alpha_1 = 0$, whereas the second half-space is
occupied by a material substance that has trivial dielectric and
magnetic properties, $\varepsilon_2 = \mu_2 = 1$, but has a constant
axion piece with $\alpha_2 = \alpha \neq 0$, see Fig.2. Let us assume,
for simplicity, that the incident wave is linearly polarized with the
electric vector directed along the $y$ axis, which is achieved by
putting $b_1=b_2=0$:
\begin{equation}
E^{\rm incident} = W_y(\xi)\,dy = (a_1\cos\xi + a_2\sin\xi)\,dy,\qquad 
\xi = \omega (t - x/c).
\end{equation}
Then we find that the reflected wave is also a linearly polarized
wave but with the electric vector tilted with respect to the y-axis
[$\eta = \omega (t + x/c)$],
\begin{equation}
  E^{\rm reflected} = R_y(\eta)\,dy + R_z(\eta)\,dz = -\,{\frac
    {\alpha/\lambda_0} {4 + \alpha^2/\lambda_0^2}}\,(a_1\cos\eta +
  a_2\sin\eta)\,\left[( \alpha/\lambda_0)\,dy + 2dz\right],
\end{equation}
whereas the transmitted wave is linearly polarized with the electric
vector also tilted with respect to both $y$ and $z$ axes,
\begin{equation}
E^{\rm transmitted} = T_y\,dy + T_z\,dz = {\frac 2 {4 + \alpha^2/\lambda_0^2}}
\,(a_1\cos\xi + a_2\sin\xi)\left[ 2dy - (\alpha/\lambda_0)\,dz\right]. 
\end{equation}
As we see, when $\alpha = 0$ the reflected wave is absent and the
incident wave propagates from vacuum into vacuum without being
distorted, $E^{\rm incident} = E^{\rm transmitted}$.  However, when
$\alpha \neq 0$ the reflection takes place! Its presence is direct
observational evidence for the axion piece. It can be used for the
experimental determination of the value of $\alpha$. A qualitative
check of the axionic nature of the substance is provided by the fact
that the polarization of the reflected wave should be rotated with
respect to the polarization of the incident wave. Furthermore, the
quantitative estimate of $\alpha$ can then be extracted from the
measurement of the intensity and the angle of rotation of the
reflected wave that explicitly depend on the value of the axion.

\section{The general case: Oblique incidence}

The above analysis can be generalized to the case when the wave is not
normally incident on the interface between the two media, see Fig.3.
Then, for an arbitrarily moving wave, we have in the first medium
(left half-space) a superposition of the incident and the reflected
waves:
\begin{eqnarray}
  E &=& \left(W_y + R_y\right)dy + W_zdz_{(i)} +
  R_zdz_{(r)},\label{E1a}\\ B &=& {\frac {k_1}
    \omega}\,\left[dx_{(i)}\wedge \left(W_y dy + W_z dz_{(i)} \right)
    - dx_{(r)}\wedge\left( R_y dy +
      R_zdz_{(r)}\right)\right],\label{B1a}\\ {\cal D} &=&
  \varepsilon_1\varepsilon_0\left[\left(W_y dz_{(i)} - W_z dy
    \right)\wedge dx_{(i)} + \left(R_y dz_{(r)} - R_z dy\right)\wedge
    dx_{(r)} \right]\nonumber\\ && +\,\alpha_1\,{\frac {k_1}
    \omega}\,\left[dx_{(i)}\wedge \left(W_y dy + W_zdz_{(i)}\right) -
    dx_{(r)}\wedge\left(R_y dy +
      R_zdz_{(r)}\right)\right],\label{D1a}\\ {\cal H} &=&
  \varepsilon_1\varepsilon_0{\frac \omega {k_1}}\left[W_y dz_{(i)} -
    R_ydz_{(r)} - \left(W_z - R_z\right)dy\right]\nonumber\\ &&
  -\,\alpha_1 \left[\left(W_y + R_y\right)dy + W_zdz_{(i)} +
    R_zdz_{(r)} \right].\label{H1a}
\end{eqnarray}
Here $x_{(i)} = x\,\cos\theta_1 + z\,\sin\theta_1$, $z_{(i)} =
-x\,\sin\theta_1 + z\,\cos\theta_1$, $x_{(r)} = x\,\cos\theta_2 -
z\,\sin\theta_2$, $z_{(r)} = x\,\sin\theta_2 + z\,\cos\theta_2$
describe the local coordinates adapted to the incident and the
reflected wave, respectively, whereas
\begin{eqnarray}
  W_y(\xi_1) &=& a_1\cos\xi_1 + a_2\sin\xi_1,\qquad W_z(\xi_1) =
  b_1\cos\xi_1 + b_2\sin\xi_1,\\ R_y(\eta)&=& c_1\cos\eta +
  c_2\sin\eta,\quad\qquad R_z(\eta) = d_1\cos\eta + d_2\sin\eta,
\end{eqnarray}
with $\xi_1 = \omega\,t - k_1x_{(i)}$ and $\eta = \omega\,t +
k_1x_{(r)}$.  The angles $\theta_1$ and $\theta_2$ give, as usual, the
angles of the incident and the reflected waves with respect to the
normal of the interface $S$.

\begin{figure}
\epsfxsize=7truecm
\epsfbox{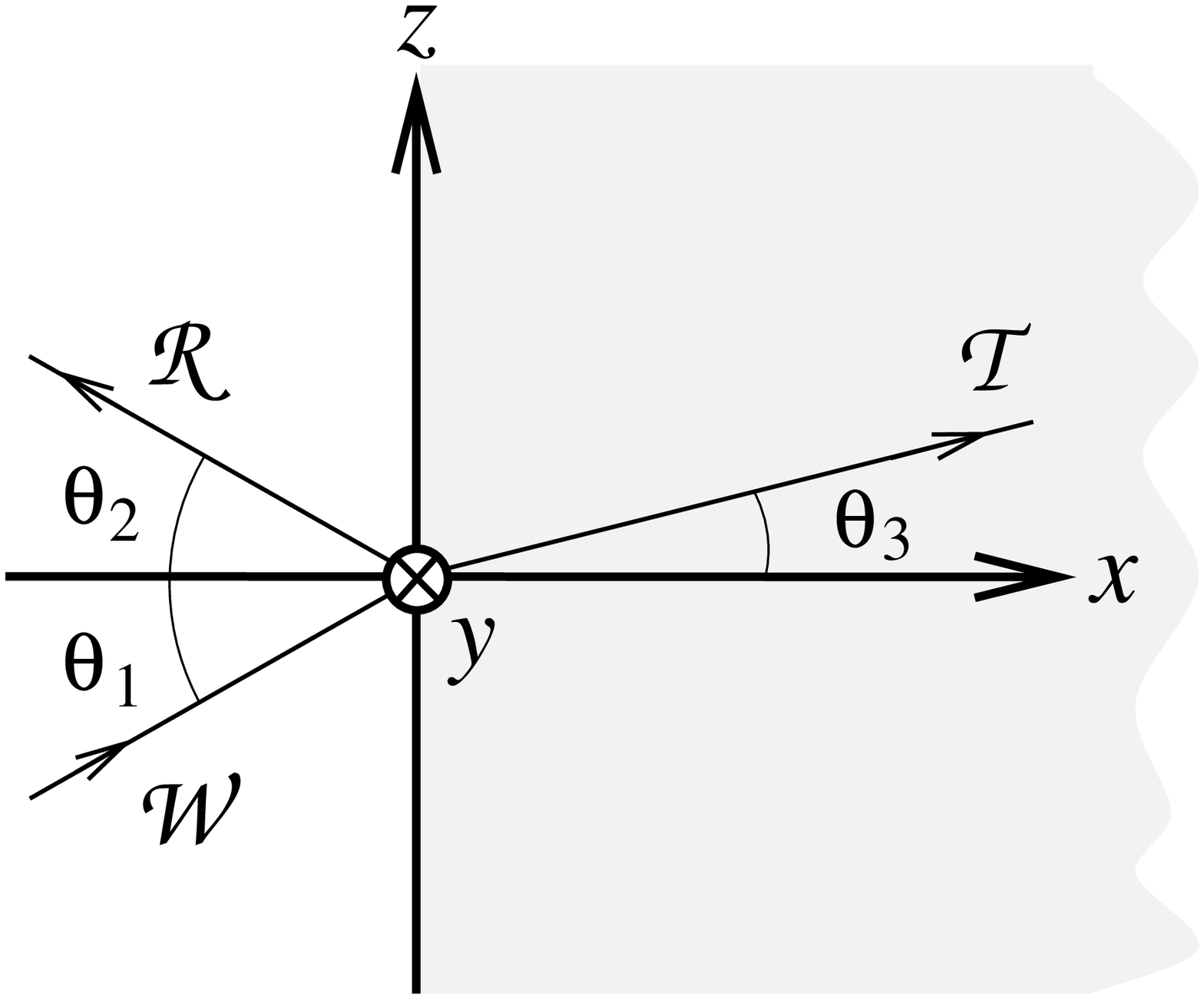}
\caption[{\it Oblique incidence. The wave\/} $\cal W$ {\it is\/} 
$\cal R${\it \!eflected and\/} $\cal T${\it \!\!ransmitted.}]
{\label{fig3}{\it Oblique incidence. The wave\/} $\cal W$ {\it is\/} 
$\cal R${\it \!eflected and\/} $\cal T${\it \!\!ransmitted.}}
\end{figure}

The transmitted wave in the right half-space has a form similar to
(\ref{E2})-(\ref{H2}):
\begin{eqnarray}
  E &=& T_y\,dy + T_z\,dz_{(t)},\label{E2a}\\ B &=& {\frac {k_2}
    \omega}\,dx_{(t)}\wedge\left(T_y\,dy +
    T_z\,dz_{(t)}\right),\label{B2a}\\ {\cal D} &=&
  \varepsilon_2\varepsilon_0\left(T_y\,dz_{(t)} - T_z\,dy\right)
  \wedge dx_{(t)} +\alpha_2\,{\frac {k_2}
    \omega}\,dx_{(t)}\wedge\left(T_y\,dy +
    T_z\,dz_{(t)}\right),\label{D2a}\\ {\cal H} &=&
  \varepsilon_2\varepsilon_0{\frac \omega {k_2}}\left(T_y\,dz_{(t)} -
    T_z\,dy\right) - \alpha_2\left(T_y\,dy +
    T_z\,dz_{(t)}\right).\label{H2a}
\end{eqnarray}
Here we denote $x_{(t)} = x\,\cos\theta_3 + z\,\sin\theta_3$, $z_{(t)}
= -x\,\sin\theta_3 + z\,\cos\theta_3$, where $\theta_3$ describes the
refraction angle. Similarly to (\ref{T1}), we have
\begin{equation}
T_y(\xi_2) = p_1\cos\xi_2 + p_2\sin\xi_2,\qquad  
T_z(\xi_2) = q_1\cos\xi_2 + q_2\sin\xi_2,\label{T1a}
\end{equation}
with the argument $\xi_2 = \omega\,t - k_2x_{(t)}$. 

Now, the jump conditions (\ref{jump1}) and (\ref{jump2}) yield
\begin{eqnarray}
  \left(W_y + R_y\right)\vline\,{\hbox{\raisebox{-1.5ex}
      {\scriptsize{x=z=0}}}} &=&
  T_y\vline\,{\hbox{\raisebox{-1.5ex}{\scriptsize{x=z=0}}}},\label{match1a}\\ 
  \left(W_z\,\cos\theta_1 + R_z\,\cos\theta_2\right)\vline
  \,{\hbox{\raisebox{-1.5ex}{\scriptsize{x=z=0}}}} &=&
  T_z\,\cos\theta_3\vline
  \,{\hbox{\raisebox{-1.5ex}{\scriptsize{x=z=0}}}},\label{match2a}\\ 
  \varepsilon_0c\sqrt{\frac
    {\varepsilon_1}{\mu_1}}\left(W_y\,\cos\theta_1 -
    R_y\,\cos\theta_2\right){}\vline\,{\hbox{\raisebox{-1.5ex}
      {\scriptsize{x=z=0}}}} &=& \left(\varepsilon_0c\sqrt{\frac
      {\varepsilon_2} {\mu_2}}\,T_y -
    [\alpha]\,T_z\right)\cos\theta_3{}\vline
  \,{\hbox{\raisebox{-1.5ex}{\scriptsize{x=z=0}}}},\label{match3a}\\ 
  \varepsilon_0c\sqrt{\frac {\varepsilon_1}{\mu_1}}\left(W_z -
    R_z\right){}
  \vline\,{\hbox{\raisebox{-1.5ex}{\scriptsize{x=z=0}}}} &=&
  \varepsilon_0c \sqrt{\frac
    {\varepsilon_2}{\mu_2}}\,T_z{}\vline\,{\hbox{\raisebox{-1.5ex}
      {\scriptsize{x=z=0}}}} +
  [\alpha]\,T_y{}\vline\,{\hbox{\raisebox{-1.5ex}
      {\scriptsize{x=z=0}}}},\label{match4a}\\ 
  k_1\left(W_y\,\sin\theta_1 + R_y\,\sin\theta_2\right){}\vline
  \,{\hbox{\raisebox{-1.5ex}{\scriptsize{x=z=0}}}} &=& k_2
  T_y\,\sin\theta_3
  \vline\,{\hbox{\raisebox{-1.5ex}{\scriptsize{x=z=0}}}},\label{match5a}\\ 
  \varepsilon_1\varepsilon_0\left(W_z\,\sin\theta_1 -
    R_z\,\sin\theta_2\right){}
  \vline\,{\hbox{\raisebox{-1.5ex}{\scriptsize{x=z=0}}}} &=&
  \left(\varepsilon_1 \varepsilon_0\,T_z + {\frac
      {[\alpha]k_2}{\omega}}\,T_y\right)\sin\theta_3
  \vline\,{\hbox{\raisebox{-1.5ex}{\scriptsize{x=z=0}}}}.\label{match6a}
\end{eqnarray}
The last two equations (when combined with the rest) yield the well
known result that relates the angles of incidence, reflection, and
refraction, namely, $\sin\theta_1 = \sin\theta_2$ (law of reflection)
and $n_1\sin\theta_1 = n_2\sin\theta_3$ (law of refraction).  In
addition, {}from the algebraic system (\ref{match1a})-(\ref{match4a}),
we find the coefficients of the reflected ($c_{1,2}$ and $d_{1,2}$)
and transmitted ($p_{1,2}$ and $q_{1,2}$) waves as combinations of
those of the incident wave:
\begin{eqnarray}
  c_{1,2} &=& {\frac {\cos\theta_3}{\Delta}}\Bigg\{\left[\left({\frac
        {\varepsilon_1} {\mu_1}} - {\frac {\varepsilon_2}{\mu_2}} -
      {\frac {[\alpha]^2}{\lambda_0^2}}\right)\cos\theta_1 +
    \sqrt{\frac {\varepsilon_1
        \varepsilon_2}{\mu_1\mu_2}}\left(\cos\theta_1 -
      \cos\theta_3\right)\right] a_{1,2}\nonumber\\ && \qquad
  +\,{\frac {[\alpha]}{\lambda_0}}\sqrt{\frac {\varepsilon_1}{\mu_1}}
  \,(\cos\theta_1 + \cos\theta_2)\,b_{1,2}\Bigg\},\\ d_{1,2} &=&
  {\frac {\cos\theta_3}{\Delta}}\Bigg[ -\,{\frac {[\alpha]}
    {\lambda_0}}\sqrt{\frac {\varepsilon_1}{\mu_1}}\,(\cos\theta_1 +
  \cos\theta_2) \,a_{1,2}\nonumber\\ &&\qquad +\,\left[\left({\frac
        {\varepsilon_1} {\mu_1}} - {\frac {\varepsilon_2}{\mu_2}} -
      {\frac {[\alpha]^2}{\lambda_0^2}}\right) \cos\theta_1 +
    \sqrt{\frac {\varepsilon_1\varepsilon_2}{\mu_1\mu_2}}
    \left(\cos\theta_3 - \cos\theta_1\right)\right] b_{1,2}\Bigg\},\\ 
  p_{1,2} &=& \sqrt{\frac {\varepsilon_1}{\mu_1}}{\frac {(\cos\theta_1
      + \cos\theta_2)}{\Delta}}\left[\left(\sqrt{\frac {\varepsilon_1}
        {\mu_1}} \,\cos\theta_3 + \sqrt{\frac
        {\varepsilon_2}{\mu_2}}\,\cos\theta_2\right) a_{1,2} + {\frac
      {[\alpha]}{\lambda_0}}\,\cos\theta_3\,b_{1,2}\right],\\ q_{1,2}
  &=& \sqrt{\frac {\varepsilon_1}{\mu_1}}{\frac {(\cos\theta_1 +
      \cos\theta_2)}{\Delta}}\left[ -\,{\frac
      {[\alpha]}{\lambda_0}}\,\cos\theta_2 \,a_{1,2} +
    \left(\sqrt{\frac {\varepsilon_1} {\mu_1}}\,\cos\theta_2 +
      \sqrt{\frac {\varepsilon_2}{\mu_2}}\,\cos\theta_3\right)
    b_{1,2}\right].
\end{eqnarray}
Here we denoted $\Delta =
\left(\cos\theta_2\,\sqrt{\varepsilon_1/\mu_1} + \cos\theta_3
  \,\sqrt{\varepsilon_2/\mu_2}\,\right)\left(\cos\theta_3\,\sqrt{
    \varepsilon_1/\mu_1} +
  \cos\theta_2\,\sqrt{\varepsilon_2/\mu_2}\,\right) +
\cos\theta_2\cos\theta_3\,[\alpha]^2/\lambda_0^2$. For the special
case $\theta_2=\theta_3=0$, we recover $\Delta_\bot$.

\section{The measureability of the axion piece}

As we saw, we can read off from the coefficients $c_{1,2}$ and
$d_{1,2}$ of the reflected wave the jump $[\a]$ of the axion piece. In
principle, one can also measure the axion piece by means of observing
the properties of the transmitted wave inside the medium. However, the
study of the reflected wave alone is clearly preferable as it provides
a simple and elegant scheme of experimental determining the properties
of a substance without destructing the latter.

\section*{Acknowledgment}

Financial support from the DFG project HE-528/20-1 is gratefully
acknowledged.

\begin{footnotesize}

\end{footnotesize}

\end{document}